\documentstyle[preprint,prl,eqsecnum,aps,graphicx]{revtex}

\begin{document}
\tightenlines
\preprint{}

\title{New Limit for the Half-Life of 2K(2$\nu$)-Capture\\
Decay Mode of $^{78}$Kr.}
\author{Ju.M. Gavriljuk$^1$, V.V. Kuzminov$^1$,
N.Ya. Osetrova$^1$, S.S. Ratkevich$^{2}$}

\address{$^1$Institute for Nuclear Research, Moscow}
\address{$^2$Kharkov National University, Ukraine}

\date{\today}

\maketitle

\begin{abstract}

Features of data accumulated at 1817 hours in the experimental
search for 2K(2$\nu$)-capture decay mode of $^{78}$Kr are
discussed. The new limit for this decay half-life is found to
be T$_{1/2}\geq 2.3 \times 10^{20}$ yr. (90$\%$ C.L.).

\end{abstract}


     First result for 2K(2$\nu$)-capture decay mode of $^{78}$Kr was
presented in refs.\cite{Gav97}-\cite{Gav98}. The limit derived
from the data collected at 254.2 hours was T$_{1/2}\geq 0.9 \times
10^{20}$ yr. (90$\%$ C.L.). The theoretical predictions for the
half-life of $^{78}$Kr(2e,2$\nu$)$^{78}$Se capture are $3.7 \times
10^{21}$ yr. \cite{Suh96}, $3.7 \times 10^{22}$ yr.
\cite{Hirsch94} and $6.2 \times 10^{23}$ yr. \cite{Urin98}.
Corresponding values of half-life for 2K(2$\nu$)-capture decay
mode are $4.7 \times 10^{21}$ yr., $4.7 \times 10^{22}$ yr. and
$7.9 \times 10^{23}$ yr. if one takes into account that
2K-electron capture part is 78.6$\%$ from a total number of
2$e$-captures for $^{78}$Kr \cite{Kotani92}. The method used in
refs.\cite{Gav97}-\cite{Gav98} allows to reach a level of
sensitivity for a half-life up to 10$^{22}$ yr. and tests some of
theoretical models. Results of the next step of measurement are
presented.

\vskip 0.6 cm

     Measurements were performed with use of the multiwire wall-less
proportional counter (MWPC) with a krypton sample enriched in
$^{78}$Kr. The main features of the counter and measurement
conditions have been described in refs.\cite{Gav97}-\cite{Gav98}.
The MWPC contents a central main counter (MC) and a surrounding it
protection ring counter (RC) in the same body. A common anode
wires signal ($PAC$) from RC and $PC1$ and $PC2$ signals from both
ends of the MC anode are read out from MWPC. A scheme with signal
read out from two sides of the MC anode allows to determine the
event relative coordinate $\beta$ along the anode ($\beta=100
\times PC1/(PC1+PC2)$) and to eliminate the events which don't
correspond to a selected working length. A shaping amplifier with
26 $\mu$s integration and differentiation shaping times was used
for the amplification of the $PC1$ and $PC2$ pulses to have a good
enough energy resolution. A parameter $f=1000 \times
P12/(PC1+PC2)$ was used to obtain an information about a pulse
rise time and the pulse front features. Signals $P12$ are output
pulses of the additional shaping amplifier which amplifies a sum
signal ($PC1+PC2$) with the 1.5 $\mu$s shaping times. A value of
parameter $f$ depends on the energy space distribution of event in
the MC volume.

The K-shell double vacancy of daughter $^{78}$Se$^{**}$ isotope
appears as a result of $^{78}$Kr 2K(2$\nu$) $^{78}$Se capture.
A total energy released is 2K$_{ab}=25.3$ keV where K$_{ab}$ is
a couple energy of K-electrons with the Se nuclear. One can
obtain that Se$^{**}$ summary probability to emit one or two
characteristic X-rays is 0.837 in assumption that this double
vacancy deexcitation is equivalent to sum of two single
vacancies deexcitation. The characteristic X-ray
(${E_{K{\alpha}}\cong 11.2}$ keV, ${E_{K{\beta}}\cong 12.5}$
keV) has a sufficient long path length in a krypton. Two
point-like energy releases with a total energy of 2K$_{ab}$
(total energy absorption peak) will appear if X-ray will absorb
in the MC working volume. One part is the X-ray energy release
and the second one is the release of Auger electron cascade
energy accompany with the characteristic L-shell X-rays energy.
The X-rays may leave the MC volume. One- or two-point event
would be detected in this case (escape peak). All single
electron background events such as Compton electrons or inner
$\beta$-decay electrons will have one-point energy releases. A
multi-point event pulse $P12$ would represent a sequence of
short pulses with a different time overlap. A number of pulses
in the burst corresponds to a number of the local regions where
a total ionization distributed. An amplitude and duration of
each pulse in a burst depend on a local track length,
orientation and distance from the MC anode. The ADCs used to
record the $PAC$, $PC1$, $PC2$ and $P12$ signals are triggered
with the input pulse amplitude maximum. The $P12$ signal
triggering will be done for the first amplitude maximum which
corresponds to an energy released in the anode nearest local
region. Peaks corresponding to one-point amplitudes $P12$ for a
fixed event total energy appear in the event number
distribution as a function of parameter $f$ ($f$-distribution).
Events with energy released in MC only and the one in the MC
and RC simultaneously named "Type 1" and "Type 2" events,
respectively.

\vskip 0.6 cm

  A  krypton enriched up to 94$\%$ in $^{78}$Kr was used  to search for the
$^{78}$Kr(2K,2$\nu$)$^{78}$Se capture mode. It content an admixture of the
natural $\beta$-radioactive  $^{85}$Kr (T$_{1/2} = 10.7$ yr,
$E_{\beta max} = 670$ keV) with the volume activity of 0.14 Bk/{\it l}.

Measurements were done in the underground laboratory of the Baksan
Neutrino Observatory of the Institute for Nuclear Research  RAS
(Moscow) at a depth of 4900 m w.e.. The MWPC was placed in the low
background shield formed by 15 cm of lead, 8 cm of borated
polyethylene, and 11 cm of copper.

The own background of the MWPC filled up to 4.8 atm with pure
xenon without radioactive contamination was measured
preliminary. A background energy spectrum 1 collected at 973.9
h and a conveniently scaled spectrum 2 of a $^{109}$Cd source
($E_{\gamma}=88$ keV) are shown on Fig. 1. The spectra consist
of ($PC1+PC2$) signals from the type 1 events. There are a peak
at $E_{\gamma}=88$ keV and the xenon escape peak at
$E=E_{\gamma}-E_{XeK{\alpha}}=88-28.9=58.2$ keV on the curve 2.
The peak at 88 kev is not symmetrical because of the radiation
scattered in the counter wall. The energy resolution of the 88
keV $\gamma$-line is 13.7$\%$.

The background spectrum has some features.  The main peaks
correspond to the energy values of 16, 35, 50, 68, 82 and 92 keV.
In the energy regions $35\div68$ keV, the are initial peaks accompanied by the
escape peaks.
The background counting rate in the energy range $20\div100$ keV
is 91 h$^{-1}$.

Energy spectra  1 and 2 of ($PC1+PC2$) signals from type 1 and type 2
events respectively  for the $^{109}$Cd calibration source and the krypton
filling are shown on Fig.2$a$.

One can see the 88 keV peak on the spectrum 1. This spectrum
was multiplied by coefficient 0.5 for convenient comparison.
The energy resolution of this peak is 10.8$\%$. The highest
energy peak on the curve 2 is the krypton escape peak with the
energy of $E=E\gamma-E_{KrK_{\alpha}}=88-12.6=75.4$ keV. It's
appearing in the type 2 events caused by an absorption in the
RC of krypton characteristics radiation from the MC. The escape
peak on the spectrum 1 is on the left slope of the total
absorption peak. The source radiation scattered in the counter
body wall lies in this region too. $f$-Distributions
correspondent to this spectra are shown on Fig.2$b$ with the
same scaling and notation. One can see a peak on the curve 1
with a maximum at $f_{1}=166$ which corresponds to two-point
events from the total absorption peak when the
KrK$_{\alpha}$-ray ionization collected first on the MC anode.
If the photoelectron ionization collected first, the events
have a peak with a maximum at $f_{2}=920$. Calculated
$f$-values of these maximums should be equal to
$f_{1}=1000\times12.6/88=143$ and
$f_{2}=1000\times(88-12.6)/88=878$ at a calibration when the
amplitude $P12$ equal to ($PC1+PC2$) one for the single-point
events. Real values calculated from the experimental data
differ slightly from the theoretical ones. This could be
explained by nonideality of the experimental set up. A value of
$f$-parameter depends on energy also because of it. The energy
spectrum 2 on Fig.2$a$ consists of one-point events mainly and
it's $f$-distribution 2 has no multi-point peaks. The peaks at
$f=1015$ and $f=1239$ are one-point peaks for the energy region
higher than 16 keV and for the krypton 12.6 keV X-ray peak
correspondingly. The $f$-distribution $1$ has a one-point peak
with the maximum at $f=1039$.

The type 1 event background energy spectrum of the MWPC with the
krypton is shown on Fig.3$a$ (spectrum 1). It collected at 1817 h. A
counting rate is 1506 h$^{-1}$ for the energy range $20\div100$ keV.
Corresponding $\beta$- and $f$-distributions are shown on Fig.3$b$
(curve 1) and Fig.3$c$ (curve 1).
One can see peaks on the ends of  the $\beta$-distribution which caused by
events from a high energy part  of the $^{85}$Kr $\beta$-spectrum mainly
collected in an ionization mode at the end effect correction anode bulges.
To eliminate this background component, it is sufficient perform the
$\beta$-selection of events in the range $36\le\beta\le58$ (Fig.3$b$,
curve 2).
The energy spectrum 2 and $f$-distribution 2  correspond to this selection.
One can see  from the $f$-distribution 2 that the background events with
$f\le710$ suppressed mainly.
The energy spectrum of events with $36\le\beta\le58$ and
$f\le710$ is shown on Fig.3$a$ (spectrum 3).
It's shape repeats main features of the spectrum 1 on the Fig.1
 with excluded escape peaks.
It means that almost all one-point events from the $^{85}$Kr
$\beta$-spectrum excluded by the used selection. The curve 5 on
the Fig.3$c$ shows roughly a shape and a place of a
$f$-distribution waited for the $^{78}$Kr(2K,2$\nu$)$^{78}$Se
multi-point events.

The events with $36\le\beta\le58$ and  $330\le f \le 710$ were
used for the final analysis because of there is no any peak-like
distortions of the residual energy spectrum $4$ (Fig.3$a$) in the
region of interest for the such selection.

A low energy part of this spectrum is shown on the Fig.4 (spectrum
$1$). A sample of a shape and a place  of the
$^{78}$Kr(2K,2$\nu$)$^{78}$Se effect is shown as a spectrum 3
(Fig.4). An energy region $25.3\pm3.8$ keV include 95$\%$ of the
events. A background was fitted by using of points before and
above this region (spectrum 2). The sum fitted background for the
$25.3\pm3.8$ keV was found to be 266. The experimental data sum is
262. The difference is $-4\pm23$ or $-19\pm111$ yr$^{-1}$. Taking
into account the efficiency of the events registration (0.22) and
the effective counter length (0.6 of working  length), we find
that the limit of the half-life of $^{78}$Kr with respect to the
2K(2$\nu$)-capture mode is T$_{1/2}\ge2.3\times10^{20}$ yr (at a
90$\%$ C.L.).

\vskip 0.4 cm

This work  was supported by Russian Foundation for Basic Research
(project nos 94-02-05954a and 97-02-16052) and in part by the
International Science Foundation (grants nos RNT000 and RNT300).

\newpage

\newpage
\begin{figure}
\begin{center}
\includegraphics[clip,width=4.000 in]{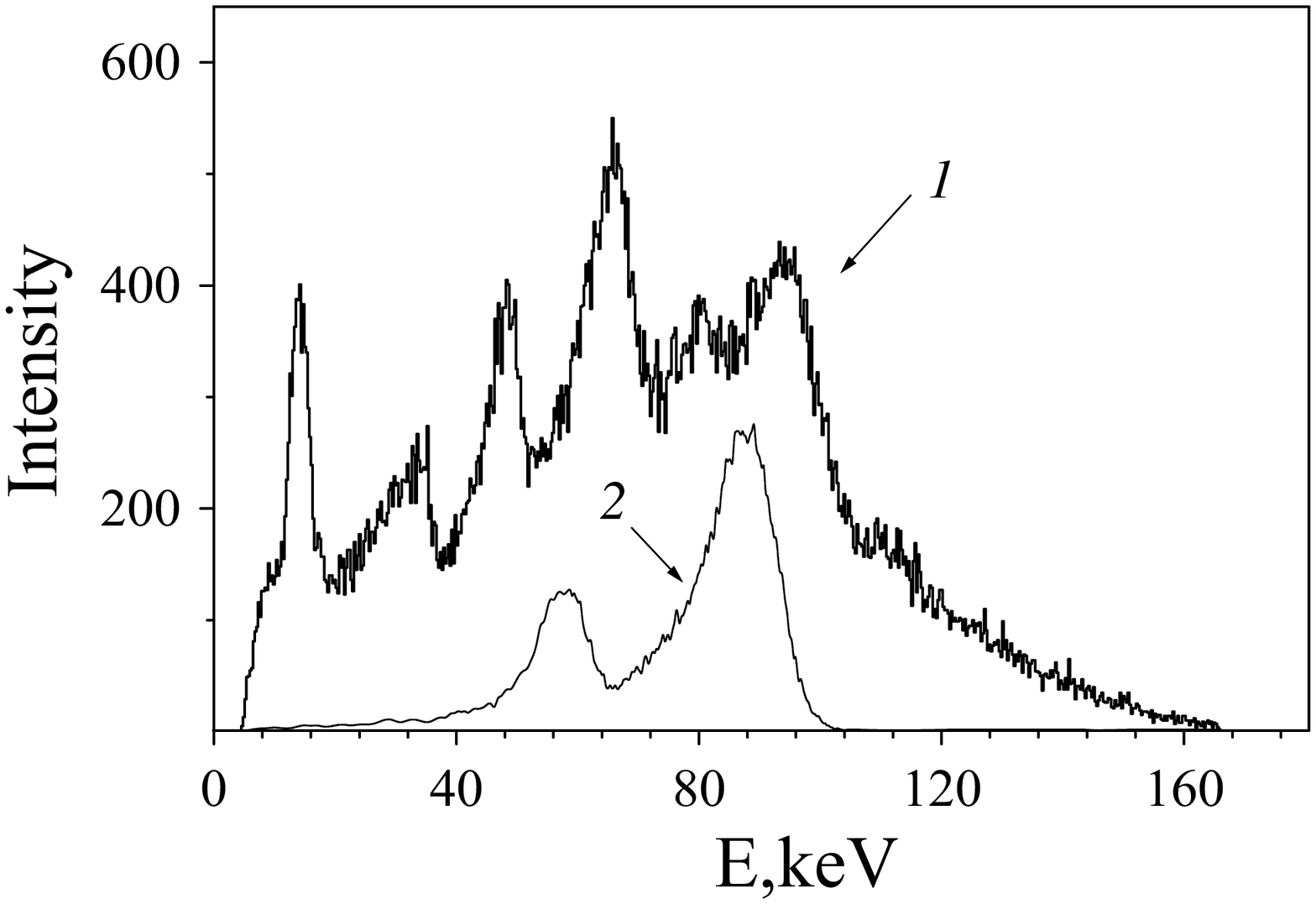}
\vskip 0.5 cm \caption{Energy spectra   of  (1)  background  and (2) $^{109}$Cd
source of  ($PC1+PC2$) signals  for type 1 events from the MWPC
filled up to 4.8 atm with pure xenon.} \label{Fig.1}
\end{center}
\end{figure}
%
\begin{figure}
\begin{center}
\includegraphics[width=4.0in]{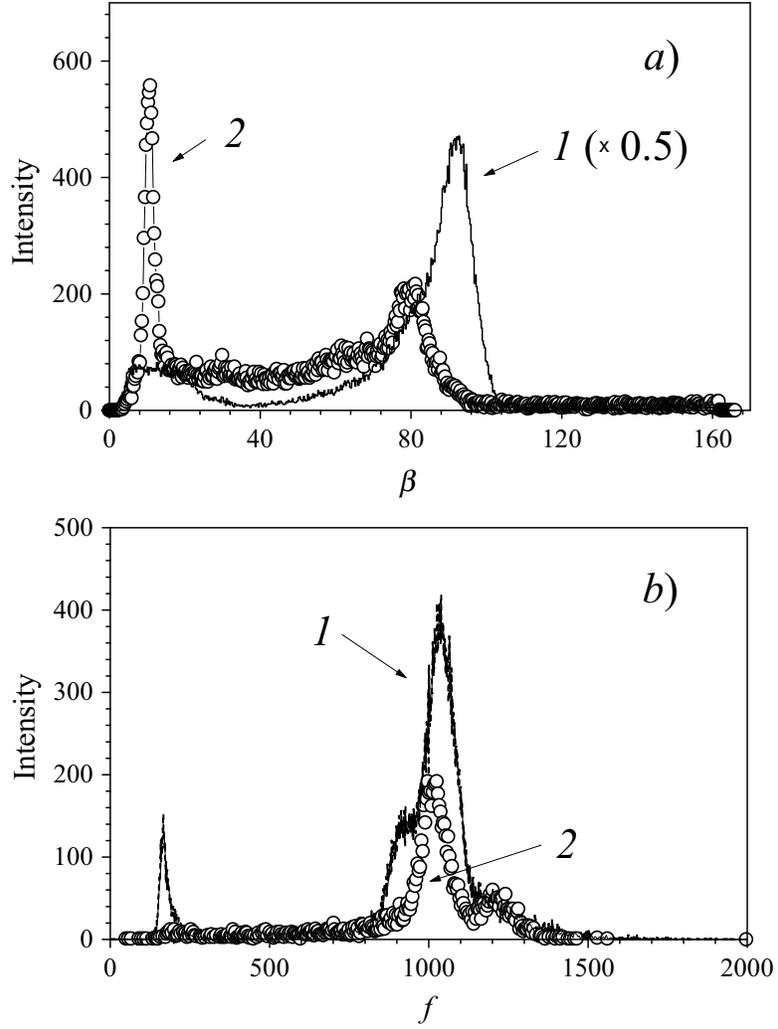}
\vskip 0.5 cm \caption{($a$) - Energy spectra of ($PC1+PC2$) signals for type 1
events ($1$) and type 2 events (2) for $^{109}$Cd source and  the
MWPC  krypton filling. ($b$) - Corresponding  $f$-distributions
(1) and (2). The curves (1)  multiplied by the coefficient 0.5.}
\label{Fig.2.}
\end{center}
\end{figure}
%
\begin{figure}
\begin{center}
\includegraphics[width=4.0 in]{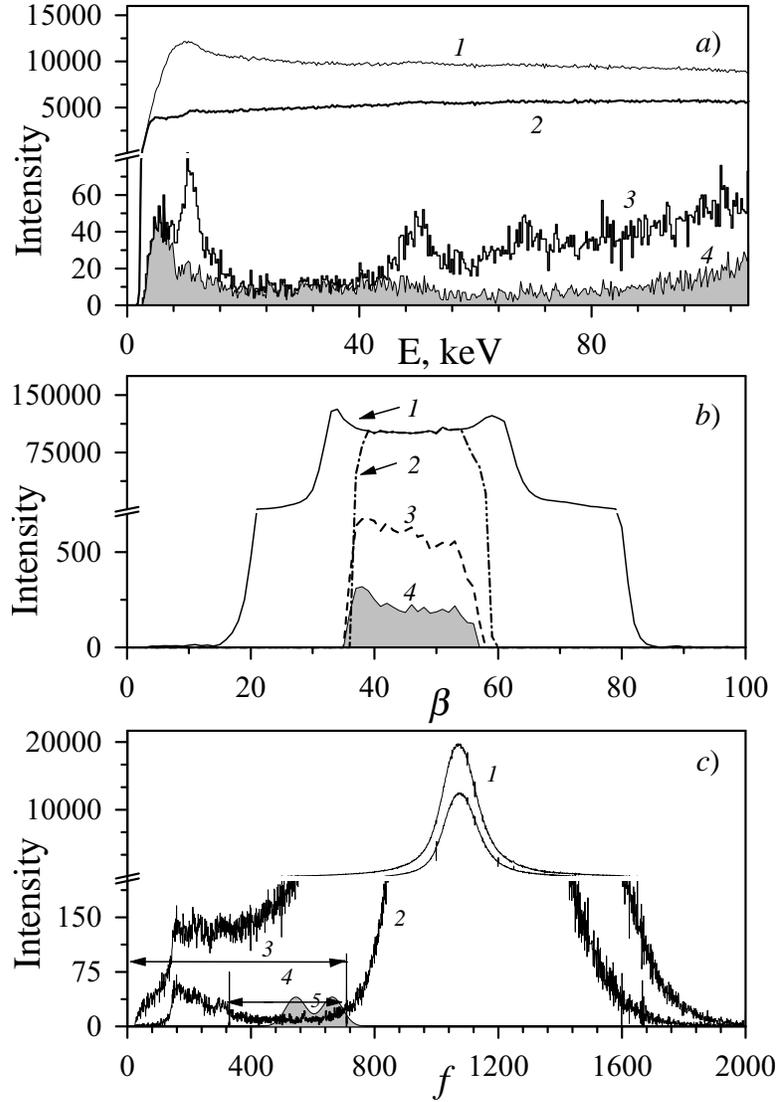}
\vskip 0.5 cm \caption{ $a$) - Background energy spectra.
$b$) and $c$) - corresponding $\beta$- and $f$-distributions.
(1) - $0\le\beta\le100$, $1\le f \le2000$; (2) -
$36\le\beta\le58$, $1\le f \le2000$;(3) - $36\le\beta\le58$,
$1\le f \le710 $; (4) - $36\le\beta\le58$, $330\le f \le710$;
(5) - sample of a shape of the $f$-distribution for the
2K-capture $^{78}$Kr events.}
\label{Fig.3.}
\end{center}
\end{figure}
\begin{figure}
\begin{center}
\includegraphics[clip,width=2.75 in]{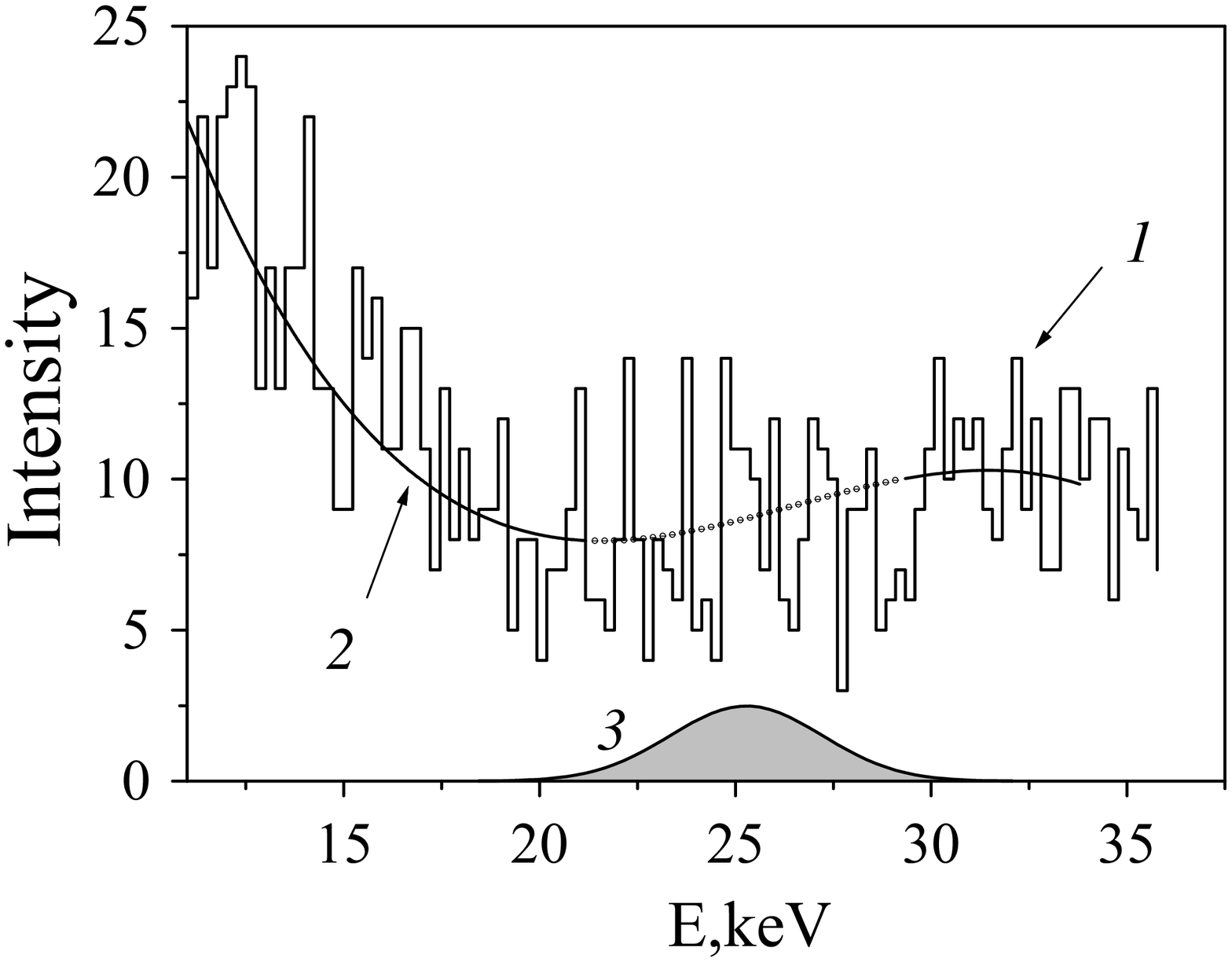}
\vskip 0.5 cm \caption{(1) - Background residual energy spectrum. (2) - The best
fit. (3) - Sample spectrum of an effect.} \label{Fig.4.}
\end{center}
\end{figure}


\begin{references}

\bibitem{Gav97} Gavriljuk, Ju. {\it et al., Proc. of the Ninth Int. School
"Particles and Cosmology"},Baksan Valley, Kabardino-Balkaria,
Russia, April 15-22, 1997.Published by the Institute for Nuclear
Research of RAS,Moscow, 1998,  p. 415.

\bibitem{Gav98} Gavriljuk,\,Ju. et al., Phys. of Atomic Nuclei,
1998, vol.{\bf 61}, p.1287.

\bibitem{Suh96} Aunola,\,M. and  Suchonen,\,J., Nucl. Phys. A,
1996, vol.602, p.133. {\bf 4}.
\bibitem{Hirsch94} Hirsch,\,M. et al., Z. Phys. A: At. Nucl., 1994, vol.
{\bf 347}, p. 151.

\bibitem{Urin98} Rumyantsev,\,O. and Urin,\,M., Phys. Lett. B, 1998, vol.
{\bf 443}, p. 51.
\bibitem{Kotani92} Doi,\,M. and Kotani,\,T., {\it Prog. Theor.
Phys.}, 1992, vol. {\bf 87}, p. 1207.
\end{references}
\end{document}